\documentclass{elsart}
\usepackage{epsf}

\begin{document}

\begin{frontmatter}

\title{Distribution of Dangling Ends on the Incipient Percolation Cluster}

\author[Giessen,Ramat-Gan]{Markus~Porto},
\author[Giessen]{Armin~Bunde}, and
\author[Ramat-Gan]{Shlomo~Havlin}
\address[Giessen]{Institut~f\"ur~Theoretische~Physik~III,
                  Justus-Liebig-Universit\"at~Giessen,
                  Heinrich-Buff-Ring~16, 35392~Giessen, Germany}
\address[Ramat-Gan]{Minerva~Center~and~Department~of~Physics,
                    Bar-Ilan~University,
                    52900~Ramat-Gan, Israel}


\begin{abstract}

We study numerically and by scaling arguments the probability $P(M) \;
\mathrm{d}M$ that a given dangling end of the incipient percolation cluster has
a mass between $M$ and $M + \mathrm{d}M$. We find by scaling arguments that
$P(M)$ decays with a power law, $P(M) \sim M^{-(1+\kappa)}$, with an exponent
$\kappa = d_f^{\mathrm{B}}/d_f$, where $d_f$ and $d_f^{\mathrm{B}}$ are the
fractal dimensions of the cluster and its backbone, respectively. Our numerical
results yield $\kappa = 0.83$ in $d=2$ and $\kappa = 0.74$ in $d=3$ in very
good agreement with theory.

\end{abstract}

\end{frontmatter}

Percolation is a standard model for structural disordered systems, its
applications range from amorphous and porous media and composites to branched
polymers, gels, and complex ionic conductors
\cite{Bunde/Havlin:1996+Sahimi:1993+Stauffer/Aharony:1992}. The origin of
anomalous diffusion in the incipient percolation cluster (i.e.\ in structural
disordered system) is still challenging and its analytical treatment is still
lacking. Its complexity arises from different contribution from the backbone
and the dangling ends of the cluster, which both slow down, in a self-similar
fashion, the motion of a random walker. The backbone of a cluster is defined as
all sites carrying current when a voltage difference is applied between two
arbitrary sites, the dangling ends are the remaining sites, each one singly
connected to the backbone. Considerably effort was spent to map percolation
cluster to comb-like structures, see e.g.\ \cite{Havlin/BenAvraham:1987} for an
overview. To find suitable simple models and to understand anomalous diffusion
better, a detailed knowledge about the internal structure of the percolation
cluster is required, including distribution of masses and lengths of dangling
ends, that represents the teeth in the comb-like structures.

In this paper we concentrate on the incipient percolation cluster at
$p_{\mathrm{c}}$, which is known to be self-similar on all length scales. The
mass $S$ within Euclidean distance $r$ scales as $S \sim r^{d_f}$, where $d_f$
is the fractal dimension. A second useful metric is the so-called `chemical'
distance $\ell$, which is defined as the length of the shortest path on the
structure. This chemical distance $\ell$ scales as $\ell \sim
r^{d_{\mathrm{min}}}$, where $d_{\mathrm{min}}$ is the fractal dimension of the
shortest path. As a result, the mass within chemical distance $\ell$ scales as
$S \sim \ell^{d_{\ell}}$, with $d_{\ell} = d_f/d_{\mathrm{min}}$. The dangling
ends are known to have the same fractal dimension as the cluster, being larger
than the fractal dimension $d_f^{\mathrm{B}}$ of the backbone (and consequently
$d_{\ell}^{\mathrm{B}} = d_f^{\mathrm{B}}/d_{\mathrm{min}} < d_{\ell})$.

To study the distributions numerically we generated large critical percolation
cluster in $d=2$ and $d=3$ using the well known Leath algorithm
\cite{Leath:1976+Alexandrowicz:1980}. The maximum size of the cluster is
$\ell_{\mathrm{max}} = 2000$ in $d=2$ ($p_{\mathrm{c}} \cong 0.5927460\ldots$)
and $\ell_{\mathrm{max}} = 1000$ in $d=3$ ($p_{\mathrm{c}} \cong
0.311606\ldots$). For each of the about $10^5$ configurations the backbone is
extracted using the improved `burning' algorithm
\cite{Porto/Bunde/Havlin/Roman:1997} (for the original burning algorithm see
\cite{Herrmann/Hong/Stanley:1984}). Then the mass $M$ and the chemical size $L$
of all dangling ends are determined, as well as their total number $N$. The
chemical size $L$ is defined as the maximum chemical distance available on the
dangling end, measured from the point where it is connected to the backbone.

We are interested in the distributions $P(M)$ and $P(L)$ of dangling ends.
Here $P(M) \; \mathrm{d}M$ gives the probability that a given dangling end has
a mass between $M$ and $M+\mathrm{d}M$, which we expect to behave as
\begin{equation}
P(M) \sim M^{-(1+\kappa)} \quad,
\end{equation}
with an exponent $\kappa$ to be determined. Accordingly, $P(L) \; \mathrm{d}L$
is the probability that the dangling end has a chemical size between $L$ and
$L+\mathrm{d}L$. Since $M(L) \sim L^{d_{\ell}}$, both quantities are related by
\begin{equation}
P(L) \propto P(M) \frac{\mathrm{d}M}{\mathrm{d}L} \sim
L^{-(1+\kappa) d_{\ell} + (d_{\ell}-1)} \quad.
\end{equation}

\begin{figure}[t]
\parbox[t]{84.5mm}{
\vspace*{1mm}
\unitlength 1.0mm
\def\epsfsize#1#2{0.72#1}
\begin{picture}(84,55)
\put(11,-6){\epsfbox{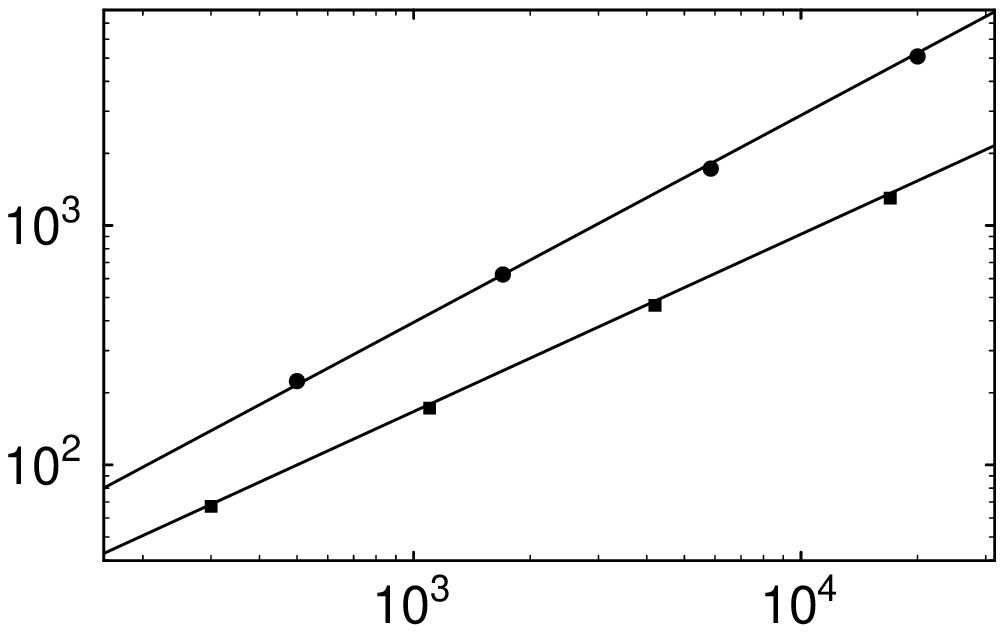}}
\put(47,-3){\makebox(10,8)[t]{\fontsize{12}{12}\selectfont $\displaystyle S$}}
\put(0,28){\makebox(10,10)[r]{\fontsize{12}{12}\selectfont $\displaystyle \frac{N(S)}{N(1)}$}}
\end{picture}}\hspace*{2mm}
\parbox[t]{50mm}{
\caption{Plot of the mean number $N(S)$ of dangling ends on a cluster of mass
$S$ in the cases $d=2$ (full circle) and $d=3$ (full square). The lines have
the slope $0.86$ ($d=2$) and $0.74$ ($d=3$).}
\label{figure:number}}
\end{figure}

For relating the exponent $\kappa$ to known exponents, let us first consider
the total number of dangling ends $N$ on a certain cluster. It is natural to
assume that this total number increases linear with the mass of the cluster's
backbone $S_{\mathrm{B}}$, i.e. $N \sim S_{\mathrm{B}}$. Since the backbone
mass scales with the cluster mass $S$ as $S_{\mathrm{B}} \sim
S^{d_f^{\mathrm{B}}/d_f}$, we obtain for the total number of dangling ends
$N(S) \sim S^{d_f^{\mathrm{B}}/d_f}$ as a function of the cluster mass $S$.
Numerical results for this quantity are shown in Fig.~\ref{figure:number}. The
obtained slopes $0.86$ ($d=2$) and $0.74$ ($d=3$) fit perfect with the ratio
$d_f^{\mathrm{B}}/d_f$. Hence, $d_f^{\mathrm{B}}$ is a very good approximation
for the fractal dimension of the number of dangling ends, i.e.\ $N \sim
r^{d_f^{\mathrm{B}}}$.

Let us further assume that the main contribution to the cluster mass $S$ is
given by the sum of the $N$ dangling end masses $M_i$ and that it is possible,
in good approximation, to neglect the backbone mass. Under this assumption we
can write approximately $S \cong \sum_{i=1}^N M_i$. When rewriting this sum by
an integral using the distribution $P(M)$, we get $S \sim N
\int_1^{M_{\mathrm{max}}} M P(M) \; \mathrm{d}M \sim N
M_{\mathrm{max}}^{1-\kappa}$, where $M_{\mathrm{max}}$ is the maximum dangling
end mass that appear in average on a cluster of mass $S$. When taking $N$
dangling end masses out of the distribution $P(M) \sim M^{-(1+\kappa)}$, the
largest one in average will be $M_{\mathrm{max}} \sim N^{1/\kappa}$. Putting
both together we get $S \sim N N^{(1-\kappa)/\kappa} \sim N^{1/\kappa}$ or $N
\sim S^{\kappa}$. Comparing this with the result $N \sim
S^{d_f^{\mathrm{B}}/d_f}$ obtained above, it follows that $\kappa =
d_f^{\mathrm{B}}/d_f$. We note that a more general derivation relating fractal
dimension and distribution was presented in \cite{Huber/Jensen/Sneppen:1995},
which yields an identical result under the same assumption that the backbone
mass is negligible compared to the cluster mass.

\begin{figure}[t]
\unitlength 1mm
\def\epsfsize#1#2{0.72#1}
\begin{center}
\begin{picture}(138,62)
\put(64,-6){\epsfbox{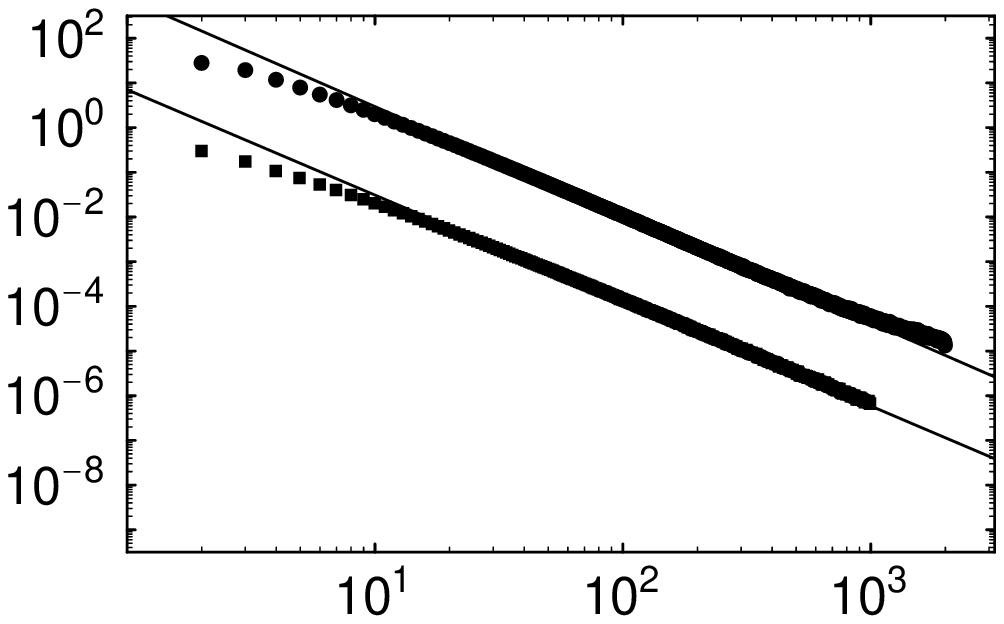}}
\put(60,10){\epsfbox{pbh350-erase.eps}}
\put(0,-6){\epsfbox{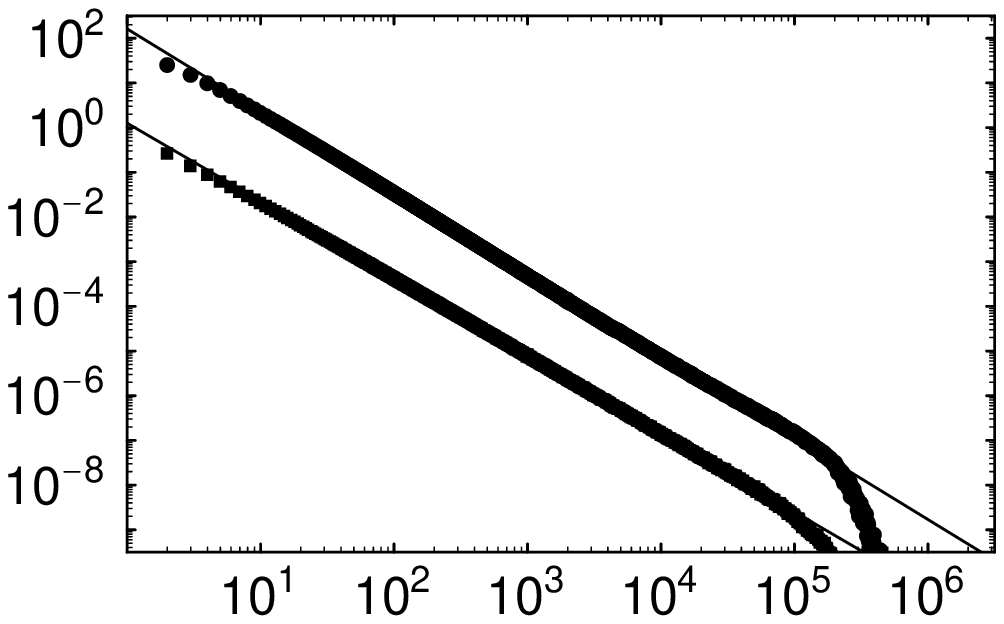}}
\put(36,-3){\makebox(10,8)[t]{\fontsize{12}{12}\selectfont $\displaystyle M$}}
\put(100,-3){\makebox(10,8)[t]{\fontsize{12}{12}\selectfont $\displaystyle L$}}
\put(36,53){\makebox(10,10)[r]{\fontsize{12}{12}\selectfont $\displaystyle P(M)$}}
\put(100,53){\makebox(10,10)[r]{\fontsize{12}{12}\selectfont $\displaystyle P(L)$}}
\put(40,32){\makebox(10,10)[r]{\fontsize{12}{12}\selectfont $\displaystyle {} \times 10^2$}}
\put(115,35){\makebox(10,10)[r]{\fontsize{12}{12}\selectfont $\displaystyle {} \times 10^2$}}
\put(60,42){\makebox(10,10)[r]{\fontsize{12}{12}\selectfont (a)}}
\put(124,42){\makebox(10,10)[r]{\fontsize{12}{12}\selectfont (b)}}
\end{picture}
\end{center}
\caption{Plot of the probability distribution functions (a)~$P(M)$ vs mass $M$
and (b)~$P(L)$ vs size $L$, in both cases for $d=2$ (full circle, multiplied by
$10^2$) and $d=3$ (full square). The lines have the slope (a)~$-1.83$ ($d=2$)
and $-1.74$ ($d=3$), and (b)~$-2.42$ ($d=2$) and $-2.36$ ($d=3$).}
\label{figure:mass+size}
\end{figure}

The results of the numerical simulation for $P(M)$ and $P(L)$ are shown in
Fig.~\ref{figure:mass+size}. For $P(M)$, we obtain a slope of $-1.83$ in $d=2$
and $-1.74$ in $d=3$, which has to be compared to $-(1+d_f^{\mathrm{B}}/d_f)$
being $-1.87$ in $d=2$ and $-1.74$ in $d=3$. Concerning $P(L)$, we find a slope
of $-2.42$ in $d=2$ and $-2.36$ in $d=3$, which has to be compared to
$-(1+d_f^{\mathrm{B}}/d_f) d_{\ell} + (d_{\ell}-1) =
-(1+d_{\ell}^{\mathrm{B}})$ being $-2.44$ in $d=2$ and $-2.36$ in $d=3$. As
conclusion, we see that $\kappa = d_f^{\mathrm{B}}/d_f$ is a very good
approximation in $d=2$ and fits even better in $d=3$. For the critical dimension
$d=6$ we recover the known result $P(M) \sim M^{-3/2}$
\cite{Bunde/Havlin:1996+Sahimi:1993+Stauffer/Aharony:1992}.

Concerning the problem of diffusion on percolation cluster and its mapping to
comb-like structures, we see that the distribution of masses $P(M)$ of dangling
ends is singular. A random comb with such a singular distribution of teeth
lengths yields anomalous diffusion \cite{Havlin/BenAvraham:1987}. Different
from this, the non-singular distribution of lengths $P(L)$ of dangling ends may
be the key quantity in the case of strong topological bias, since we expect
there that only the chemical size of the dangling ends controls the time the
random walker spends inside, and not their mass nor their internal structure.

This work has been supported by the German-Israeli Foundation, the Minerva
Center for the Physics of Mesoscopics, Fractals and Neural Networks; the
Alexander-von-Hum\-bolt Foundation; and the Deut\-sche
For\-schungs\-gemein\-schaft.

\end{document}